\documentclass[final,3p,times,twocolumn]{elsarticle}

\usepackage{amssymb}
\usepackage{array}
\usepackage{makecell}
\usepackage{fullpage}

\usepackage{color,graphicx,hyperref,amsmath}
\usepackage{mathtools}
\usepackage{graphicx}
\usepackage{float}
\usepackage{xcolor}

\journal{Chaos, Solitons \& Fractals}

\begin{document}

\begin{frontmatter}



\title{Frustration Induced Chimeras and Motion in Two Dimensional Swarmalators}


\author[inst1]{R. Senthamizhan}
\author[inst1]{R. Gopal}
\ead{gopalphysics@gmail.com}
\author[inst1]{V. K. Chandrasekar}
\ead{chandru25nld@gmail.com}

\affiliation[inst1]{organization={Department of Physics},
            addressline={Centre for Nonlinear Science and Engineering, School of Electrical and Electronics Engineering,SASTRA Deemed University}, 
            city={Thanjavur},
            postcode={613 401}, 
            state={Tamil Nadu},
            country={India}
            }

\begin{abstract}
Swarmalators are phase oscillators capable of simultaneous swarming and synchronization, making them potential candidates for replicating complex dynamical states. In this work, we explore the effects of a frustration parameter in the phase interaction functions of a two-dimensional swarmalator model inspired by the solvable Sakaguchi-swarmalators that move in a one-dimensional ring. The impact of the frustration parameter in these models has been a topic of great interest. Real-world coupled systems with frustration exhibit remarkable collective dynamical states, underscoring the relevance of this study. The frustration parameter induces various states exhibiting non-stationarity, chimeric clustering, and global translational motion, where swarmalators move spontaneously in two-dimensional space. We investigate the characteristics of these states and their responses to changes in the frustration parameter. Notably, the emergence of chimeric states suggests the crucial role of non-stationarity in phase interactions for spontaneous population clustering. Additionally, we examine how phase non-stationarity influences the spatial positions of swarmalators and provide a classification of these states based on different order parameters.
\end{abstract}




\end{frontmatter}



\section{Introduction}

 Synchronization is a captivating phenomenon prevalent in nature and various artificial systems \cite{bayani}. It manifests in diverse forms, from the firing of neurons \cite{penn} and the flashing of fireflies \cite{mccrea} to the oscillations of metronomes on a shared platform \cite{pantaleone}. Arthur Winfree introduced the first significant mathematical model to study synchronization in 1967 \cite{winfree}. Winfree's model was groundbreaking and paved the way for further research. Building on this, Yoshiki Kuramoto presented a simplified and analytically tractable model in 1975 \cite{kuramoto}. The Kuramoto model has since become a cornerstone in understanding synchronization, extensively studied in systems such as power systems analysis \cite{guo}, ecological systems \cite{vandermeer}, and chemical systems \cite{kuramoto_2}, providing profound insights into the underlying mechanisms of collective behavior.

Building on the understanding of synchronization, another intriguing phenomenon, swarming, involves the spatial aggregation of multiple agents \cite{schranz}. While synchronization demonstrates self-organization in time, swarming shows self-organization in space. A notable model for swarming was introduced by Vicsek in 1995 \cite{vicsek}. The Vicsek model provides the base for understanding collective motion and behavior in systems of self-propelled particles, serving as a fundamental framework in studying swarm dynamics, \cite{lu} and active matter physics \cite{puzzo}. Also, this model has been instrumental in understanding behaviors in natural systems like bird flocking, fish schooling, etc. \cite{christo, npetal}. Despite being studied independently, efforts have been made to unify synchronization and swarming. For instance, models such as mobile oscillators and chemotactic oscillators have emerged, combining these phenomena \cite{meli, fujiwara, sawai}.
 
A foundational work to combine these phenomena was conducted by Tanaka and Iwasa, who introduced a generalized model of chemotactic oscillators capable of exhibiting diverse dynamical behaviors \cite{tanaka, iwasa}. Building on this, a significant advancement in the field is the ``swarmalator" model introduced by O'Keeffe, Hong, and Strogatz in 2017 \cite{okeeffe}, which integrates synchronization and swarming behaviors within a unified framework, allowing the study of systems where individuals exhibit both spatial clustering and phase coherence. This model applies to various natural systems, including Japanese tree frogs, sperms, and vinegar eels \cite{aihara, riedel, peshkov}. Since its introduction, the swarmalator model has been explored in different configurations, considering the effects of noise, delay, pinning, forcing, higher-order interactions, Higher-harmonic interactions, and time-varying interactions \cite{hong, blum, sar1, sar2, lizarraga, anwar, smith, senthamizhan, sar3}.

Since the two-dimensional swarmalator model introduced by O'Keeffe et al. exhibits a chaotic nature \cite{ansarinasab}, a solvable model of swarmalators, which includes a pair of Kuramoto models, has also been developed \cite{yoon} and studied under various settings \cite{okeeffe2, sar4}. An exciting modification involves including a frustration parameter or phase lag in the phase interaction function. Sakaguchi and Kuramoto extensively studied the impact of this frustration parameter in the Kuramoto model in 1986 \cite{sakaguchi}. Recently, the study of frustration parameters has also been incorporated into the solvable model of swarmalators by Lizárraga et al., \cite{lizarraga1}, identifying several interesting states, such as the active asynchronous and turbulent states. 

The effect of the frustration parameter in the Kuramoto model has been studied extensively under various settings \cite{mano, mano1}, revealing rich dynamics including the emergence of frequency synchronization \cite{hsia}, complex phase locking patterns \cite{moyal}, and the presence of chimera states \cite{marten} where coherent and incoherent oscillations coexist within the same system.One notable outcome is the induction of frustration, which prevents the oscillators from achieving a perfect synchronization \cite{ha,arenas}. Our research explores the introduction of phase frustration in the two-dimensional swarmalator model by adding a phase lag which act as a frustration parameter in the phase interaction functions. Since the phase is coupled to the spatial position in our model, this non-stationarity results in unique states ranging from radially symmetric chimera to states with global translational motion. Although the two-dimensional model of swarmalators is not exactly solvable, it yields a plethora of states exhibiting interesting dynamics, paving the way for a deeper understanding of several naturally occurring swarming phenomena, such as chimeric flashing in \textit{Photuris frontalis} fireflies where both synchronized cluster and independent flashers exist in a single swarm \cite{sarfati} and swarm motility in several species of bacteria like \textit{Bacillus subtilis}, in which a multicellular group of bacteria executes a coordinated movement through the synchronization of flagella as a result of hydrodynamic interactions \cite{kearns, kearns1}. 

This paper is organized as follows: the introduction is provided in Sec. I, followed by a description of the model in Sec. II. Sec. III further explores the characteristics of chimeric, translational, and other existing states and the effect of changes in the frustration parameter on the emerging translational states. Sec. IV discusses the order parameters used and the existence of different states with respect to a range of interaction strengths. Finally, Sec. V concludes the paper.

\section{Model}
\label{sec:model}
We examine an $N$-particle model known as swarmalators in a two dimensional space $(\bf x \in \mathbb{R}^2)$ with coupled phase and spatial dynamics proposed by O'Keeffe et. al. \cite{okeeffe} to investigate various collective dynamical states. In this study, we consider a generalized swarmalator model, incorporating frustration parameters as given in the Sakaguchi-Kuramoto model \cite{sakaguchi}, described by the equations ($\dot{\bf x}$, $\dot{\theta}$) as follows,

\begin{align}
	\dot{{\bf x}}_i &= v_i + \frac{1}{N}\left[\sum_{j \neq i}^{N} \frac{{\bf x}_j-{\bf x}_i}{\left| {\bf x}_j-{\bf x}_i \right|^{\epsilon_1}}(A+J\cos(\theta_j-\theta_i+\alpha_{1})) \right. \nonumber \\
	&\qquad \left.-B\frac{{\bf x}_j-{\bf x}_i}{\left| {\bf x}_j-{\bf x}_i \right|^{\epsilon_2}}\right], \nonumber \\
	\dot{\theta}_i &= \omega_i+\frac{K}{N}\sum_{j \neq i}^{N}\frac{\sin(\theta_j-\theta_i+\alpha_{2})}{\left|{\bf x}_j-{\bf x}_i\right|^{\epsilon_3}}.
	\label{model}
\end{align}

Where, $\bf x_i = (\it x_i, \it y_i)$ denotes the spatial coordinate vector of the $i^{th}$ particle, and $\theta_i$ denotes the internal phase of the $i^{th}$ swarmalator. $N$ is the total number of swarmalators. The parameters $v_i$ and $\omega_i$, representing the velocity and natural frequency of the $i^{th}$ swarmalator, respectively, can be set to zero by selecting an appropriate reference frame without any loss of generality. The value of $A$ is set to unity to ensure that the spatial attraction function in $\dot{{\bf x}}_i$ is positive definite, while the repulsion strength $B$ is assigned as $1$. The interaction strengths $K$ and $J$ quantify the influence of surrounding swarmalators on the phase and spatial coordinates of the $i^{th}$ swarmalator, respectively. To achieve long-range attraction and short-range repulsion among the swarmalators, the exponents are set as $\epsilon_1 = 1$ and $\epsilon_2 = 2$. Additionally, $\epsilon_3$ is set to 1 to ensure that the influence of spatial distance in the phase equation $\dot{\theta_i}$ remains long-range, consistent with the phase-dependent spatial attraction.

In this model, the parameters $\alpha_{2}$ and $\alpha_{1}$ (both $ < \frac{\pi}{2}$) represent phase shifts. The interplay between $\alpha_{1}$, $\alpha_{2}$, and the coupling constants $K$ and $J$ induces non-stationarity in the system. Extending frustration parameters to both phase interaction functions in the $xy$ model reveals multiple non-stationary states. The emerging dynamical states and the order parameters used are explored in the subsequent sections.

\section{Results and Discussion}
All the simulations were performed using the RK-4 numerical algorithm with a population size of $N=500$ and a step size of $dt=0.01$. The swarmalators were initialized with positions  $x$ and $y$ uniformly distributed in the range $(-1,1)$. Their phases $\theta$ were randomly initialized following a uniform distribution in the range $[0,2\pi)$. For various pairs of $J$ and $K$ with $\alpha_{2} = \alpha_{1} < \frac{\pi}{2}$, the swarmalator system exhibits several unique dynamical states. These include static states, such as the static asynchronous state (SAS) and static chimera (SC), as well as active states, including the active phase wave (APW), active synchronized state (AS), and active chimera (AC). Additionally, the system demonstrates globally dynamic states like global translational motion (GTM) and synchronized global translational motion (SGTM). It is worth noting that varying $\alpha_{1}$ and $\alpha_{2}$ independently near $\frac{\pi}{2}$ does not introduce new states in the $JK$ parameter space. These states are discussed in detail in the following subsections.

\subsection{Static states}
By introducing the frustration parameters $\alpha_{2}$ and $\alpha_{1}$ in Eq.(\ref{model}), two distinct static states are observed: the static asynchronous state and the static chimera state.

\begin{figure}[h]
	\centering
	\includegraphics[width=8cm]{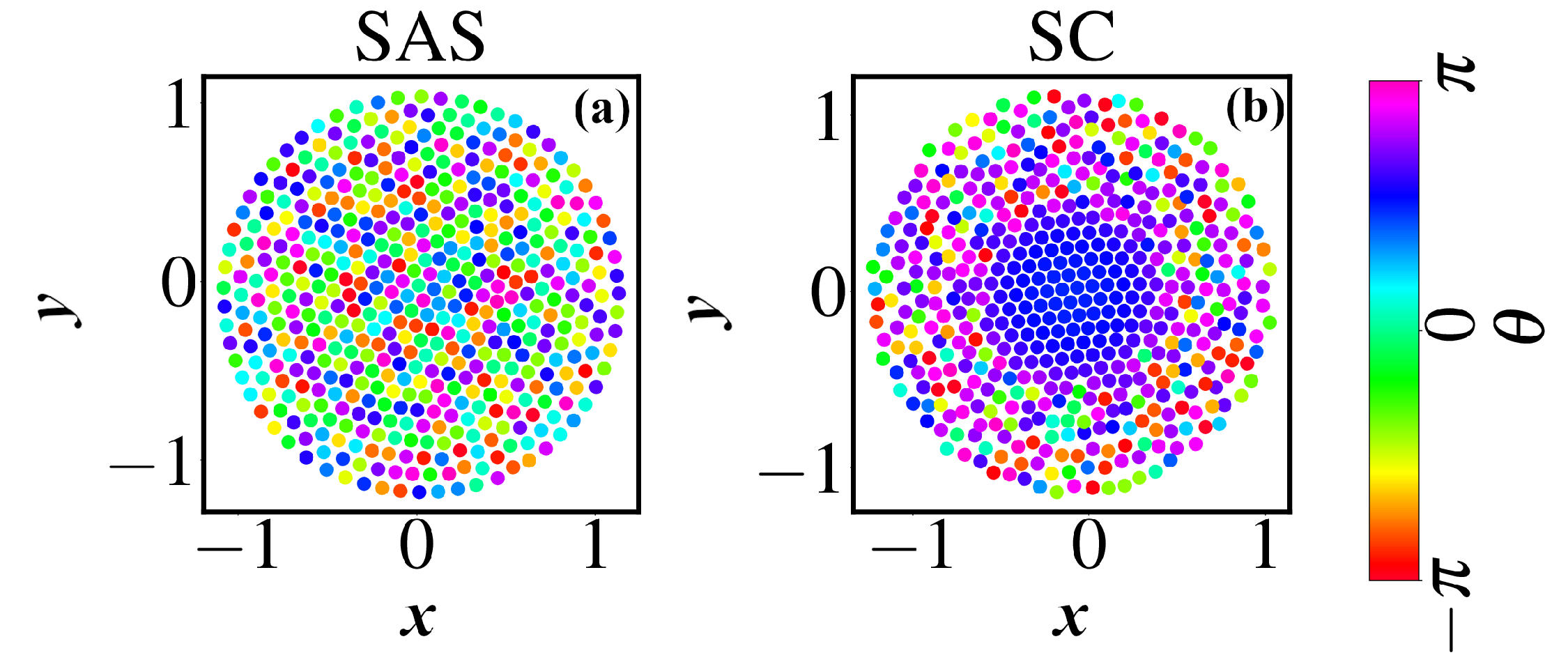}
	\caption{Static states obtained for different pairs of interaction strengths $J$ and $K$ , with $\alpha_{1} = \alpha_{2} = 1.568$}: (a) Static asynchronous state (SAS) for $J=0.05$, $K=0.05$, (b) Static chimera (SC) showing the radially symmetric clustering behavior but without any within-population motion for $J=0.05$, $K=1.0$.
	\label{fig1}
\end{figure}

Static asynchronous states (SAS) arise for small values of $ J $ and $ K $, such as $ J = 0.05 $ , $ K = 0.05 $, and $\alpha_{1} = \alpha_{2}=1.568$ as depicted in Figure \ref{fig1} (a). In this state, the swarmalators are fixed in position, forming a circular disk-like pattern. A random distribution of phases characterizes static asynchronous state. This randomness in phase distribution indicates a lack of global synchronization despite the spatial organization.

\begin{figure}[h]
	\hspace*{-0.3cm}
	\includegraphics[width=8cm]{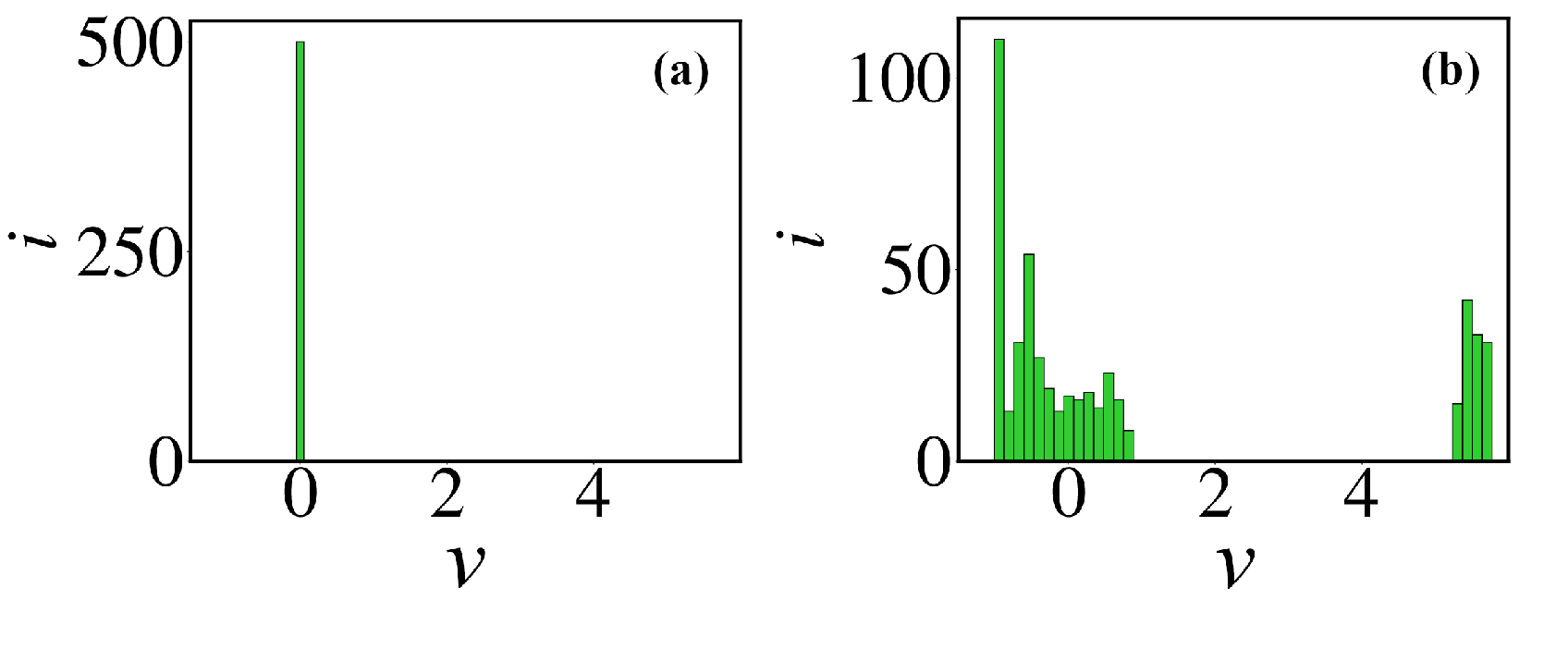}
	\caption{Histogram of phase frequency $\nu$ shows the frequency distribution of (a) static asynchronous state (SAS) with near zero frequencies and (b) static chimera state (SC) with multiple frequencies.
	}
	\label{fig2}
	\end{figure}

In the static chimera (SC) state, the swarmalators show the co-existence of synchronized and desynchronized population. Here, the synchronized population occupies the central position in the radial arrangement. In contrast, the desynchronized population forms a ring-like structure around the synchronized group, as shown in Figure \ref{fig1} (b). This formation of distinct populations occurs spontaneously, without any conditional separation of the swarmalator populations. Similar to the static asynchronous state, the phases continuously evolve due to the frustration in the phase interaction functions.   One may also note that the introduction of frustration in phase oscillators results in chimera states with clusters having locked and nonlocked frequencies in Ref.\cite{julian}. Similarly, we have also analyzed the frequency of each swarmalator with the help of the frequency histogram, shown in Figure \ref{fig2}, and it reveals the existence of distinct synchronized and desynchronized clusters within the population in SC state. 

From figure \ref{fig2} (a); it is evident that the frequency of all the swarmalators is identical in the static asynchronous state, indicating they are frequency-locked despite their randomly distributed phases. Even though the SAS shows a near-zero frequency, as shown in figure \ref{fig2} (a), it still exhibits non-stationarity, and the low values of the interaction strengths $J$ and $K$ explain the near zero frequency. In contrast, the static chimera state shows that the synchronized cluster is concentrated around the frequency $\nu = -1$ and a few randomly distributed frequencies as seen in the figure \ref{fig2} (b). This confirms that the SC state exhibits chimeric properties.  

In static chimera state, although oscillators are clustered into synchronous and asynchronous groups, the spatial distribution of these clusters exhibits radial symmetry. Notably, the synchronous cluster ceases to exist beyond a certain radius from the centroid of the population. The radius of the synchronized population $R_c$ is calculated from the time-averaged order parameter from Equation \ref{kopr} which calculates the degree of synchronization of swarmalators inside the specified radius $R$, 

\begin{align}
	\langle r(R) \rangle_t = \left\langle \frac{1}{N_R} \left| \sum_{i \in \{i: r_i \leq R\}} e^{\mathrm{i} \theta_i(R)} \right| \right\rangle_t
	\label{kopr}
\end{align}

\begin{figure}[h]
	\hspace*{-0.5cm}
	\includegraphics[width=8cm]{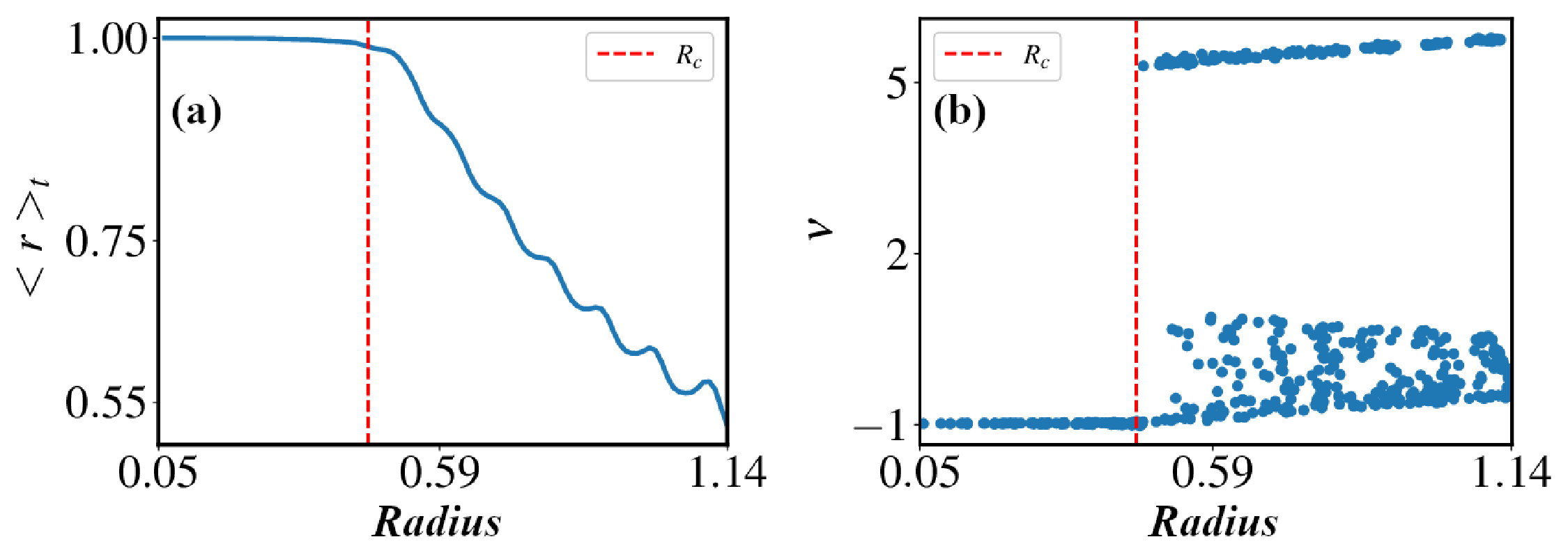}
	\caption{(a) Time averaged order parameter vs. arbitrary radius $R$ shows the order parameter from Eq. \ref{kopr}, calculated for the population $N_R$ as the radius $R$ varies. The critical radius $R_c$, beyond which the synchronized population ceases to exist, is indicated by the red dashed line. (b) Frequency vs. radius confirms that the population beyond $R_c$ is desynchronized.
	}
	\label{fig3}
\end{figure}

Where, $r_i$ is the distance of the $i^{th}$ swarmalator from the spatial center of the population, $R$ is the arbitrary radius from the spatial center of the population, $N_R$ is the number of swarmalators within the radius $R$. Here, the order parameter value remains near one when the radius includes only the synchronized population. However, as the value of that radius increases and begins to encompass the desynchronized population, the respective order parameter value decreases, as shown in Figure \ref{fig3} (a). The radius at which the order parameter starts to fall below $0.98$ is marked as the critical radius $R_c$ for the synchronized population.

Figure \ref{fig3}(b) indicates that the frequency $\nu$ of swarmalators within the critical radius $R_c$ takes a single value, whereas the desynchronized population exhibits multiple frequencies. This observation confirms the chimeric characteristics of the SC in the spatial domain.

\subsection{Active states}

So far, we have discussed scenarios where the swarmalators exhibit a static spatial property. However, the swarmalators exhibit active behavior under certain parametric conditions, where they do not remain stationary and change positions over time. The resulting active states are discussed as follows.

\begin{figure}[h]
	\hspace*{-0.3cm}
	\includegraphics[width=8cm]{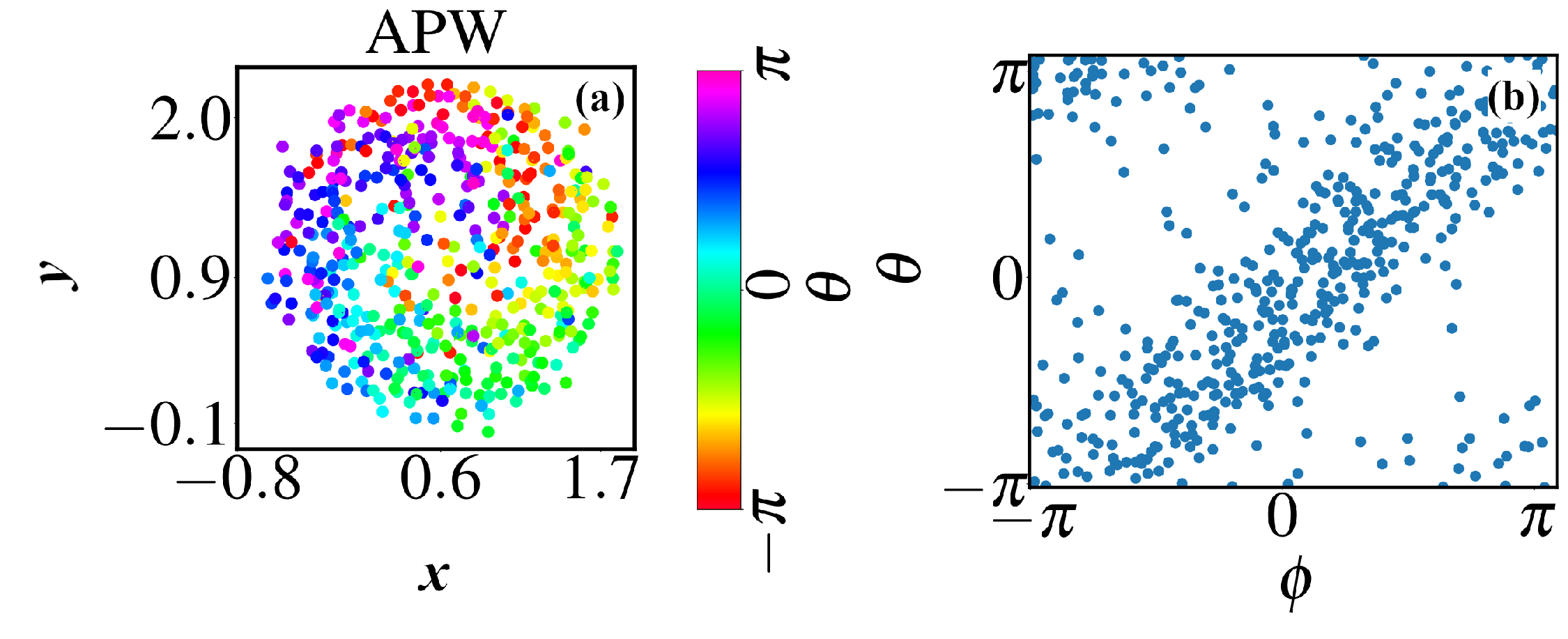}
	\caption{Snapshot of the active phase wave (APW) obtained for $J = 1.0$ , $K = 0.05$ and $\alpha_{1} = \alpha_{2}=1.568$. (b) Spatial angle $\phi_i$ vs. $\theta$ of the APW, depicting the correlation between $\theta$ and $\phi$.
	}
	\label{fig4}
\end{figure}

Due to the inclusion of the frustration parameters $\alpha_{1}$ and $\alpha_{2}$, five different active states are observed for various pairs of $J$ and $K$. One of these states is the active phase wave (APW), documented in various literature sources \cite{okeeffe, sar3}. In the APW state, the swarmalators form an annular structure and execute spatial counter-rotation within the ring structure, as shown in Figure \ref{fig4} (a). As the term `phase wave' suggests, the swarmalators' phases $\theta_i$ are correlated with their spatial angles $\phi_i$, defined as $\phi_i=\tan^{-1}\bigg(\frac{y_i-c_y}{x_i-c_x}\bigg)$, where $(c_ x,c_y)$ represents the centroid of the population, as depicted in Figure 4(b). Unlike the previously reported APW state, due to the inclusion of frustration parameters, Figure \ref{fig4} demonstrates the APW state with few number of swarmalators occupies the central region.

\begin{figure}[h]
	\centering
	\includegraphics[width=8cm]{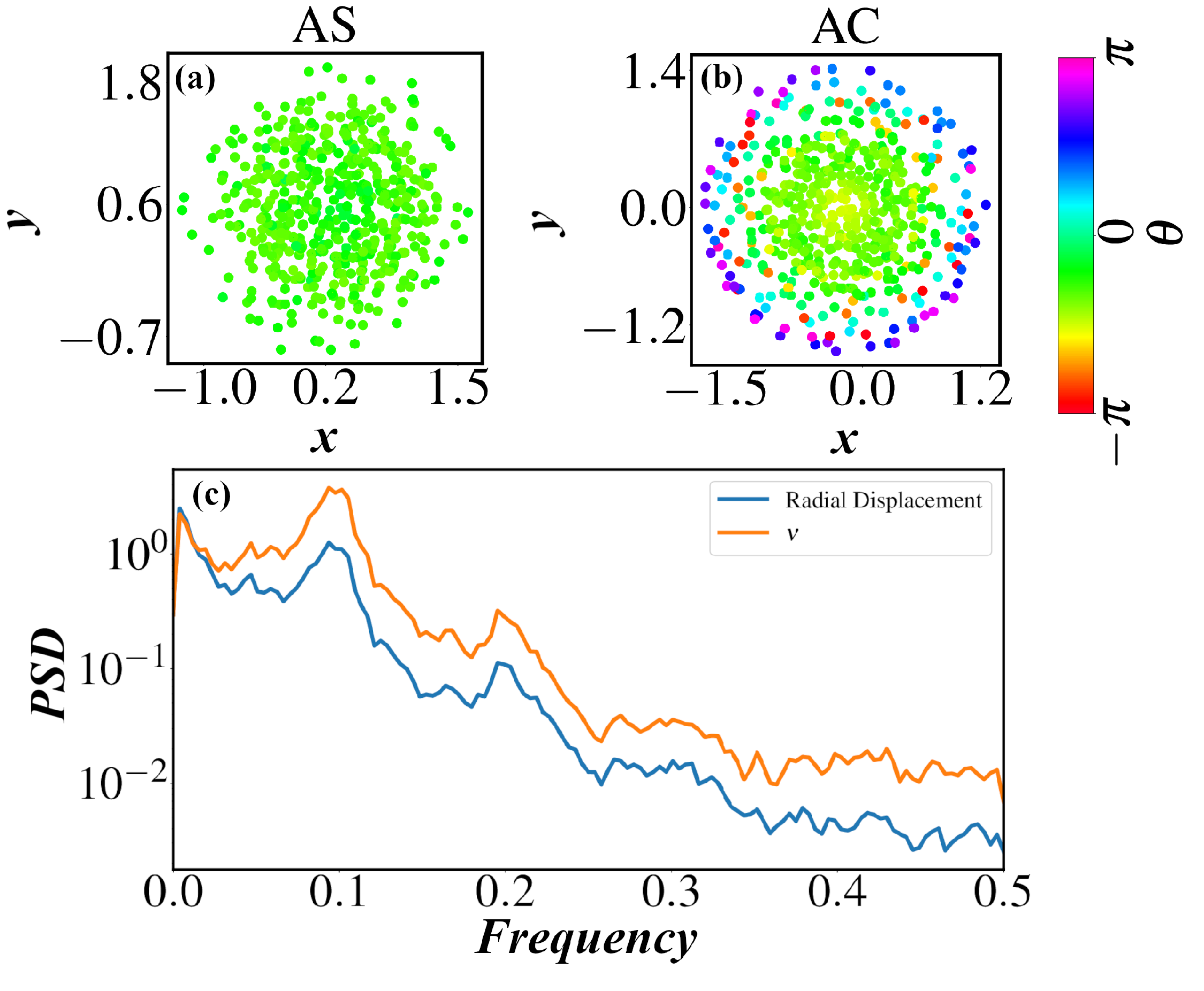}
	\caption{(a) Active synchronized (AS) state obtained for $J = 1.0$ ,$K = -0.25$, and (b) active chimera (AC) state obtained for $J = 0.55$, $K = 0.8$, (c) The power spectral density obtained from the time series of the radial displacement of the $i=100^{\text{th}}$ swarmalator and it's phase frequency shows the correlation between them and the other fixed parameters are $\alpha_{1} = \alpha_{2}=1.568$
	}
	\label{fig5}
\end{figure}

For $J = 1.0$, $K = -0.25$, and $\alpha_{1} = \alpha_{2}=1.568$, the active synchronized (AS) state is observed. In the active synchronized state (Figure \ref{fig5} (a)), the swarmalators achieve fully synchronized phases despite the negative phase coupling $K$. This phase synchronization is particularly interesting because the system’s non-stationarity favors phase synchronization mediated by the spatial attraction strength $J$, which would not be possible in the absence of frustration parameters. Also, the swarmalators execute a radial oscillation, moving from the center of the population to the outer edge and then back to the center. This movement repeats over time. The change in the radial distance of a swarmalator from the center of the population is correlated with the rate of change of $\theta$ (frequency $\nu$). This correlation is confirmed by calculating the power spectral density of the time series of both the radial displacement and the frequency of a random swarmalator, and we can confirm from figure \ref{fig5} (c) that the spectral density shows the existence of same frequency spectrum for both the time series.

\begin{figure}[h]
	\hspace*{-0.0cm}
	\includegraphics[width=8cm]{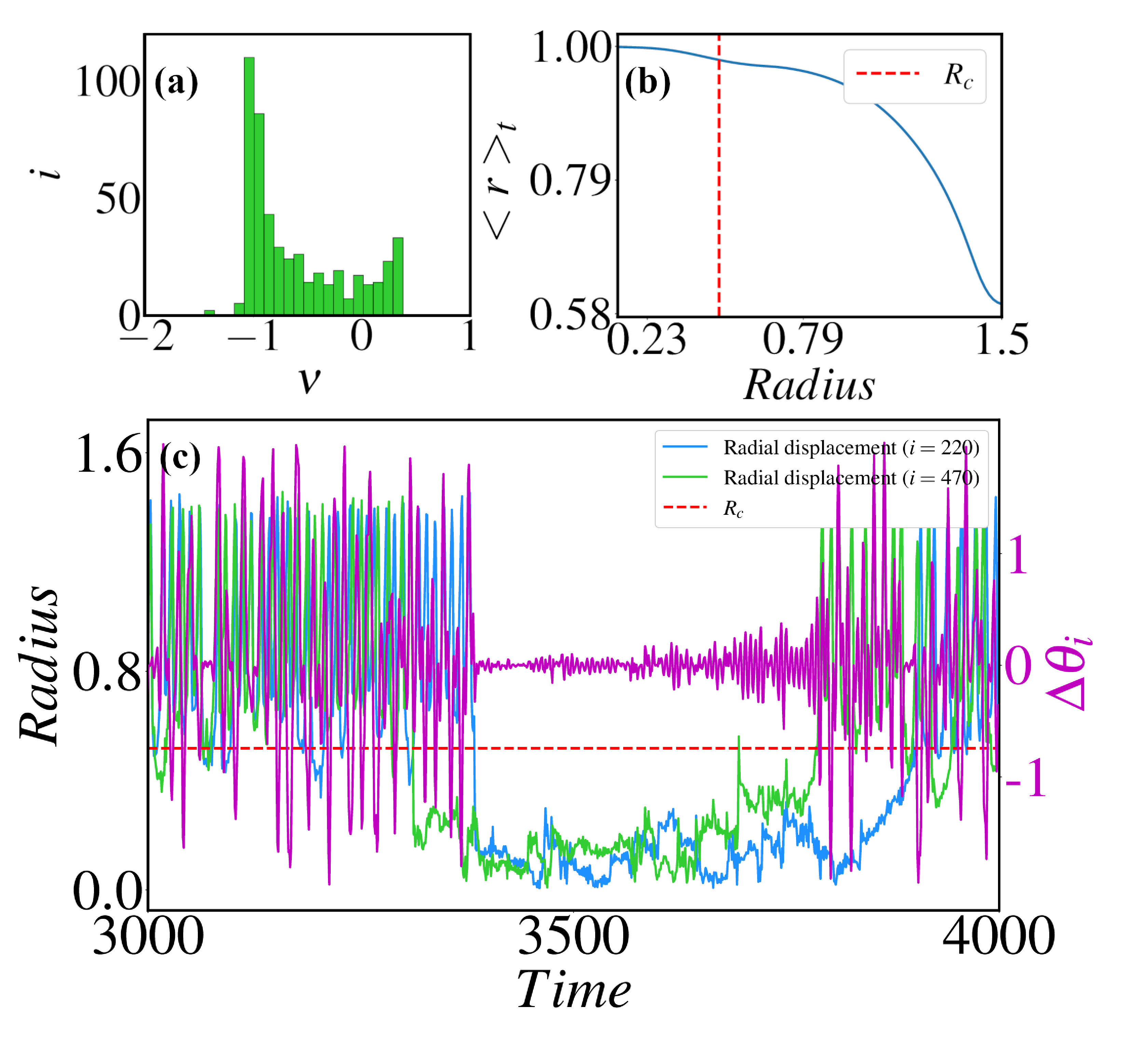}
	\caption{(a)The frequency $\nu$ versus swarmalator index $i$ histogram for the active chimera state shows the disproportionate existence of frequencies of synchronized and desynchronized populations, (b) shows the critical radius ($R_c$) at which the order parameter $<r>_t$ starts to decrease, (c) radial displacement time series shows the random sample swarmalators ($i=220,470$) shows near zero phase difference $\Delta \theta_i$ when the swarmalators move inside the critical radius $R_c$
	}
	\label{fig6}
\end{figure}

Like the static chimera state, the active chimera depicted in Figure \ref{fig5} (b) also possesses synchronized and desynchronized populations. The main difference between these states is that the swarmalators do not remain in either of the population constantly. Due to the radial oscillation of the swarmalators in the active chimera, similar to the active synchronized state, the swarmalators from the synchronized population in the central region of the disc move to the outer desynchronized ring and then fall back to the central synchronized population with the evolution of time. The number of swarmalators in the synchronized population is more significant in the active chimera state compared to the static chimera state at any instant of time. This can be verified using the frequency histogram of the AC state shown in Figure \ref{fig6} (a).

The critical radius $R_c$ calculated using the equation \ref{kopr} is marked at the radius where the order parameter value falls below the threshold value of 0.98, as depicted in Figure \ref{fig6}(b). Since there is an active exchange of swarmalators between synchronized and desynchronized populations, this active chimera state resembles a breathing chimera state observed in phase oscillators, where the synchronized and desynchronized domains oscillates in time \cite{bolotov, chenko, sheeba} leading to the continuous switching of oscillators between both domains. This behavior is confirmed by Figure \ref{fig6}(c), where the time series of radial displacement of two random swarmalators $i=220,470$ shows the switching between the synchronized and desynchronized regions in space by moving inside and out of the critical radius $R_c$. When both swarmalators fall below $R_c$, they exhibit a near-zero phase difference $\Delta \theta_i$, emphasizing synchronization and when both oscillators move above the radius $R_c$, their phase difference grows and oscillates randomly, confirming desynchronization. 

\subsection{Active states with translational motion}
\begin{figure}[h]
	\hspace*{-0.1cm}
	\includegraphics[width=8cm]{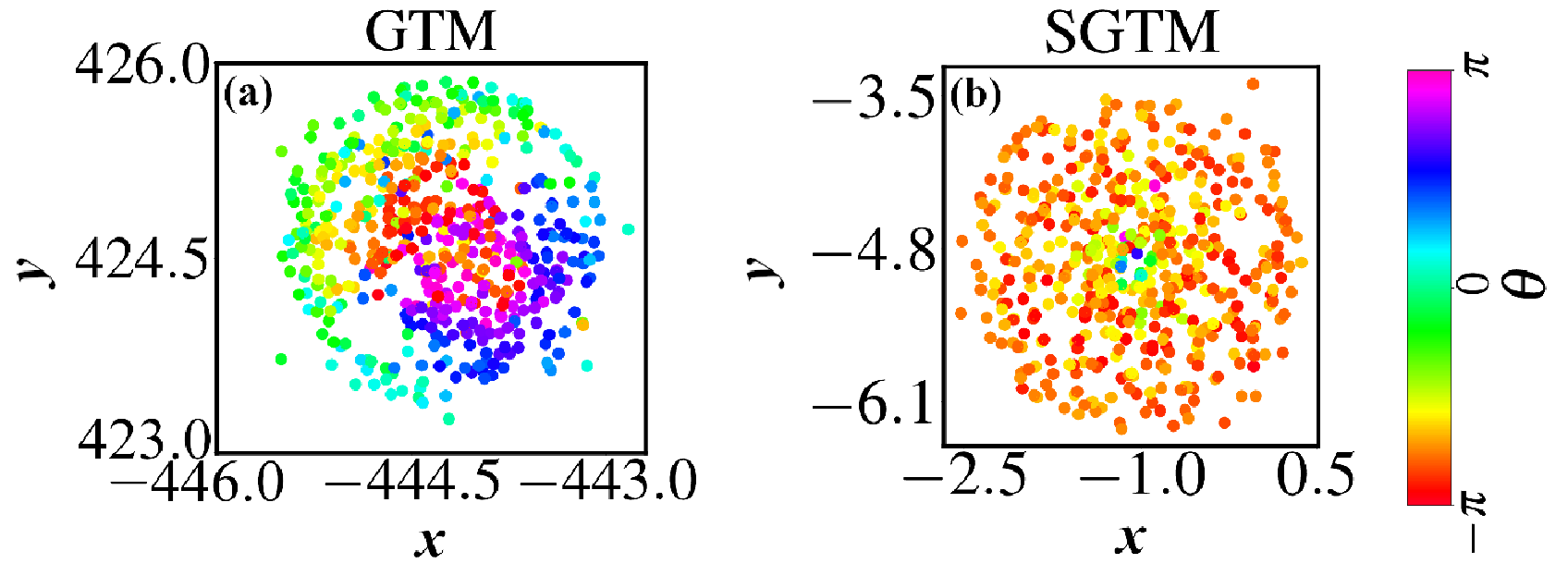}
	\caption{Snapshot of (a) global translational motion (GTM) for $J = 1.0$, $K = 0.32$, (b)synchronized global translational motion (SGTM) for $J = -1.0$ and $K = 1.0$,  and $\alpha_{1} = \alpha_{2} =1.568$ for both GTM and SGTM
	}
	\label{fig7}
\end{figure}

In this section, we discuss states with translational motion, which is characterized by the uniform collective motion of the swarm ~\cite{cavagna}, even though individual swarmalators execute motion within the population. This collective motion in $xy$ space is due to the inclusion of frustration parameters $\alpha_{1}$ and $\alpha_{2}$ (see equation \ref{model}), which induce non-stationarity in both phase and spatial position of each swarmalator. Here, we observed two states with non-stationary spatial dynamics: the global translational motion (GTM) state ($J = 1.0$ and $K = 0.32$) with wide distribution of phases and the synchronized global translational motion (SGTM) state ($J = -1.0$ and $K = 1.0$), in which the swarmalators show synchronized phases, which are shown in Figure \ref{fig7}(a) and \ref{fig7}(b), respectively.
 
Here,  in GTM, a subgroup in the central region of the spatial structure which has the highest degree of directional synchrony  aligns towards the  direction of the entire population’s global collective motion and is identified by calculating the velocity vector  $\vec{v(t)}$ of each swarmalator which is given by

 \begin{align}
 	\vec{v(t)} = \frac{\Delta \mathbf{r}}{\Delta t} = \frac{(x(t+\Delta t) - x(t)) \vec{i} + (y(t+\Delta t) - y(t)) \vec{j}}{\Delta t}
 	\label{velocity}
 \end{align}

\begin{figure}[h]
 	\hspace*{-0.1cm}
 	\includegraphics[width=8cm]{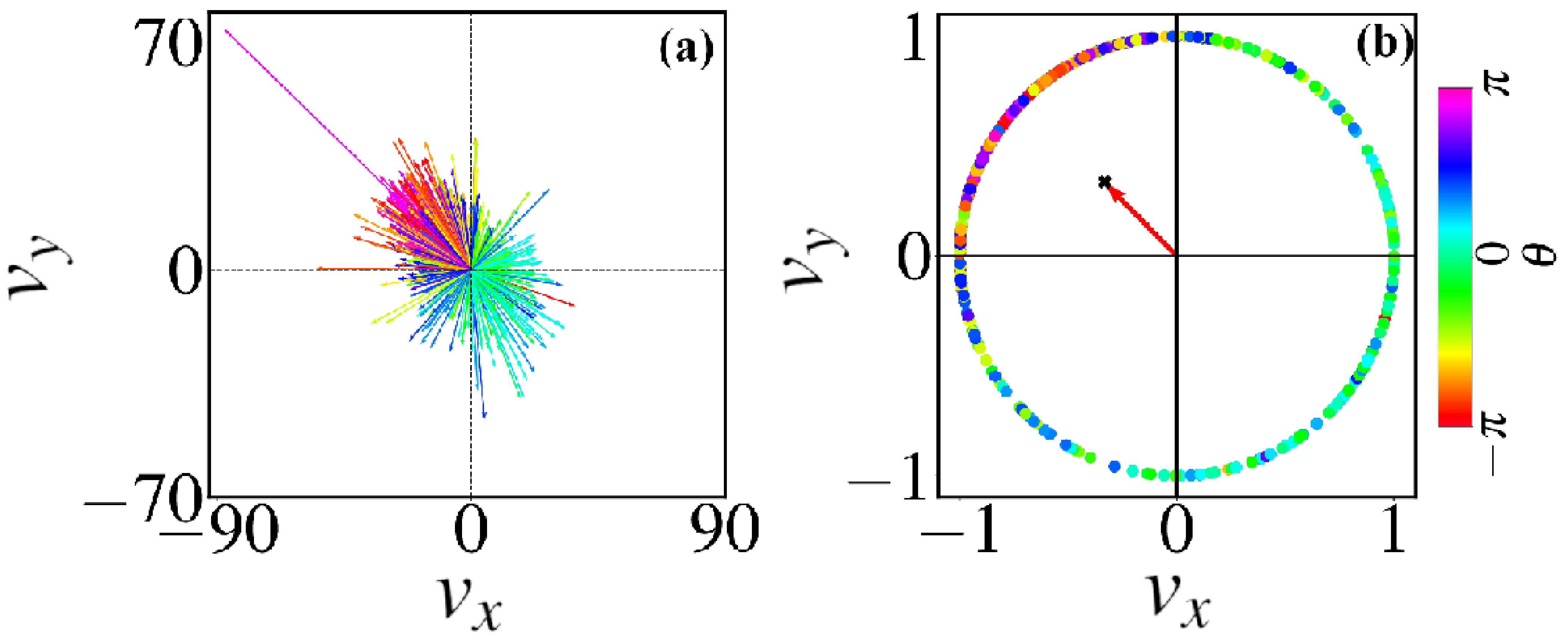}
 	\caption{(a) Velocity vectors of swarmalators in GTM translated to a common origin. (b) Normalized velocity vectors with the Kuramoto order parameter (equation \ref{dkop}) depicted as an arrow.
 	}
 	\label{fig8}
 \end{figure}

To analyze the directional properties of these velocity vectors, each vector $\vec{v(t)}$ is translated to a common origin as seen in figure \ref{fig8} (a). Translation operation ensures that the analysis is centered on the directional characteristics rather than the positions of the vectors. After translation, the vectors are normalized, resulting in vectors with their heads placed on a unit circle as given in figure \ref{fig8} (b). This transformation allows for a precise analysis of the predominant orientation of the vectors. The heads of the normalized velocity vectors are analogous to the phase oscillators in unit circle, therefore we used the Kuramoto order parameter given in the equation \ref{dkop} to determine at which direction most of the vectors are aligned, the angle of the order parameter shows the direction of alignment of the predominant velocity vectors and the magnitude, represented by the length of the arrow, indicates the degree of synchrony.

 \begin{figure}[h]
 	\hspace*{-0.3cm}
 	\includegraphics[width=8cm]{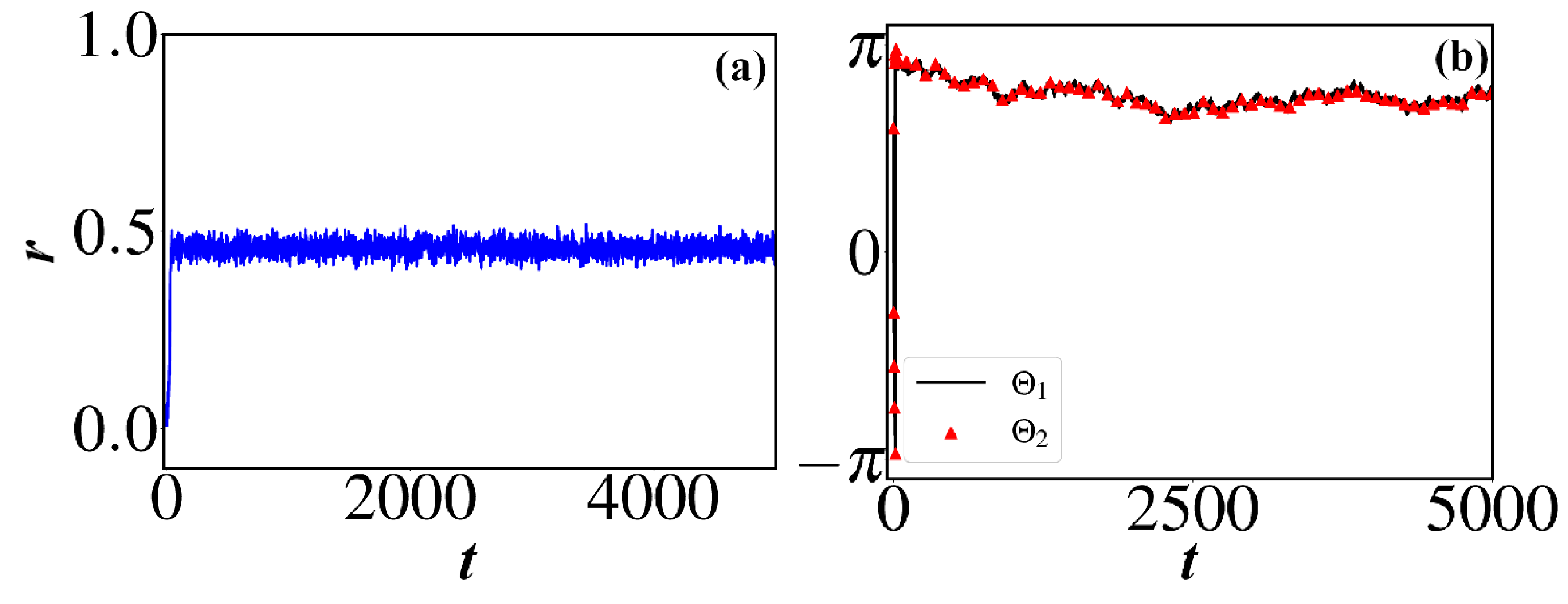}
 	\caption{(a) Time series of the magnitude of the Kuramoto order parameter obtained from the velocity vectors. (b) Time evolution of the angle $\Theta_1$ of the Kuramoto order parameter. $\Theta_2$ is the time evolution of the velocity vector obtained from the centroid of the entire population of swarmalators.
 	}
 	\label{fig9}
 \end{figure}
 
The order parameter for these velocity vectors are calculated over time and the time series of the magnitude $r$ of the Kuramoto order parameter in figure \ref{fig9} (a) demonstrates that $r \approx 0.5$ over time, indicating consistent directional synchrony within the population regardless of individual movements of the swarmalators. Additionally, the angular evolution of the Kuramoto order parameter, $\Theta_1(t)$, given by
\begin{align}
	\Theta_1(t) = \arg\left( \frac{1}{N}\sum_{j=1}^{N} e^{\mathrm{i} \psi_{j}(t)} \right),
	\label{dkop}
\end{align}
\begin{align}
	\psi_i(t) = \tan^{-1}\left(\frac{\hat{v}_y(t)_i}{\hat{v}_x(t)_i}\right)
\end{align}
 Where $\hat{v}_x(t)_i$ and $\hat{v}_y(t)_i$ are the velocity vectors corresponding to $i^{th}$ swarmalator at time $t$. Futher, $\Theta_1(t)$ compared with the angular evolution of the centroid's velocity vector, $\Theta_{2}(t)$ from equation \ref{theta2}.
 \begin{align}
 	\Theta_{2}(t) = \tan^{-1}\left(\frac{\Delta c_{y}(t)}{\Delta c_{x}(t)}\right)
 	\label{theta2}
 \end{align}
 
 Here, $\Delta c_{x}(t), \Delta c_{y}(t)$ is the velocity of centroid in $x$ and $y$ coordinates respectively at time $t$. The comparison reveals that the directional evolution of the predominant group calculated from the order parameter aligns with that of the centroid, underscoring the influence of the directionally synchronized swarmalators on the overall motion of the population as shown in figure \ref{fig9} (b). 

  The above study clearly emphasizes that the global translational motion of the swarmalators depends on the within-population directional synchrony of the swarmalator group in the central region of the spatial structure and the population in the central region is actively replaced by other swarmalators, preserving the number of directionally synchronized swarmalators. In addition to the above dynamics, unlike all other states, the spatial distribution of swarmalators in the circular disc arrangement is not uniform at any instant, as is clearly evident from Figure \ref{fig7}(a), where swarmalators form two less dense regions adjacent to the central area.

\begin{figure}[h]
	\hspace*{-0.3cm}
	\includegraphics[width=6cm]{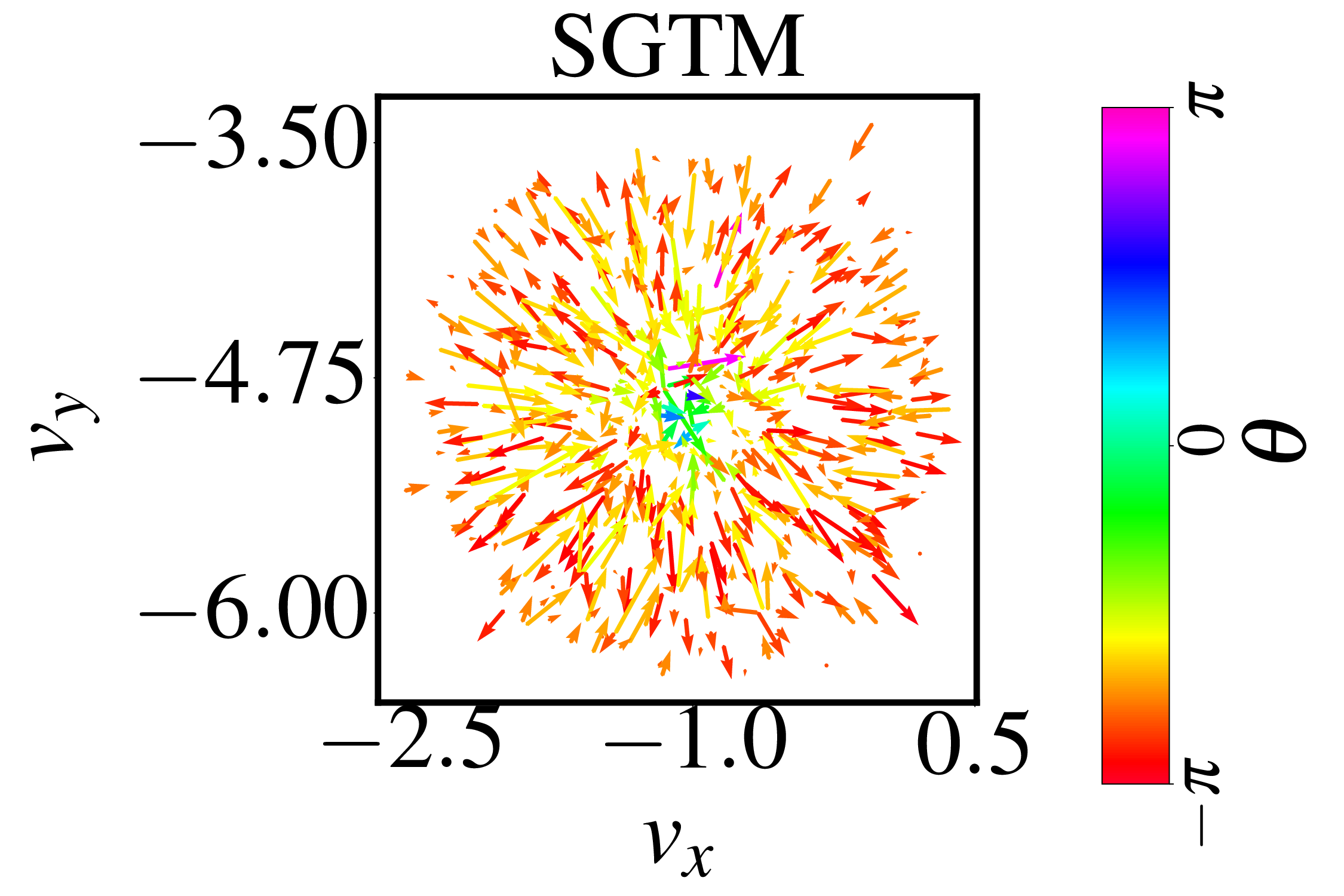}
	\caption{The velocity vectors of swarmalators in SGTM state shows the radial alignment with vectors pointing towards and away from the center. The vector's bases are drawn from the position of it's corresponding swarmalator and scaled by the factor of $0.01$ for better visualization.}
	\label{fig10}
\end{figure}

In the case of the SGTM state, the swarmalator exhibits high degree of phase synchronization when compared to GTM. A small population at the center of the radial arrangement exhibits desynchronized phases. SGTM doesn't show within population directional synchrony, because of it's radial oscillatory nature similar to the active chimera state and it is shown in Figure \ref{fig10}. 

\begin{figure}[h]
	\hspace*{-0.3cm}
	\includegraphics[width=8cm]{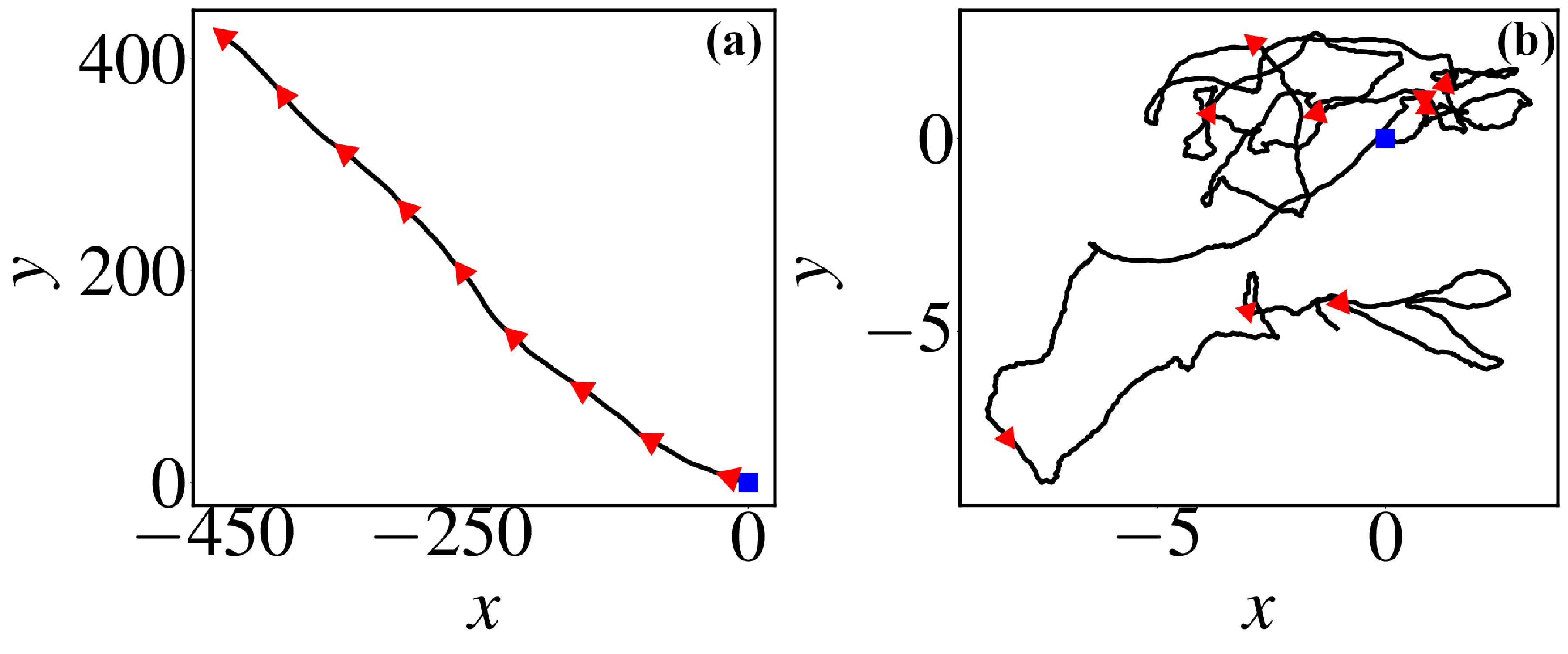}
	\caption{Evolution of the centroid over time for (a) GTM and (b) SGTM. The blue square marker indicates the initial position, and the red arrows show the direction of the centroid's evolution.
	}
	\label{fig11}
\end{figure}

We can observe that the properties of GTM and SGTM exhibit distinct characteristics and possess differing magnitudes of translational velocity. However, to compare the centroid velocity of GTM and SGTM, we plotted the displacement of the centroid in $xy$ space (see figure \ref{fig11}), in which the arrows represents the direction of movement of the centroid, and it can be observed from the figure \ref{fig11}, that the translational motion of SGTM has a relatively low centroid velocity, results in the small displacement (See Fig. \ref{fig11}(b)), compared with the GTM state, where we observe a relatively large displacement (See Fig. \ref{fig11}(a)). This emphasizes the crucial role of directional synchrony in the global spatial displacement of GTM states.

\begin{figure}[h]
	\hspace*{-0.3cm}
	\includegraphics[width=6cm]{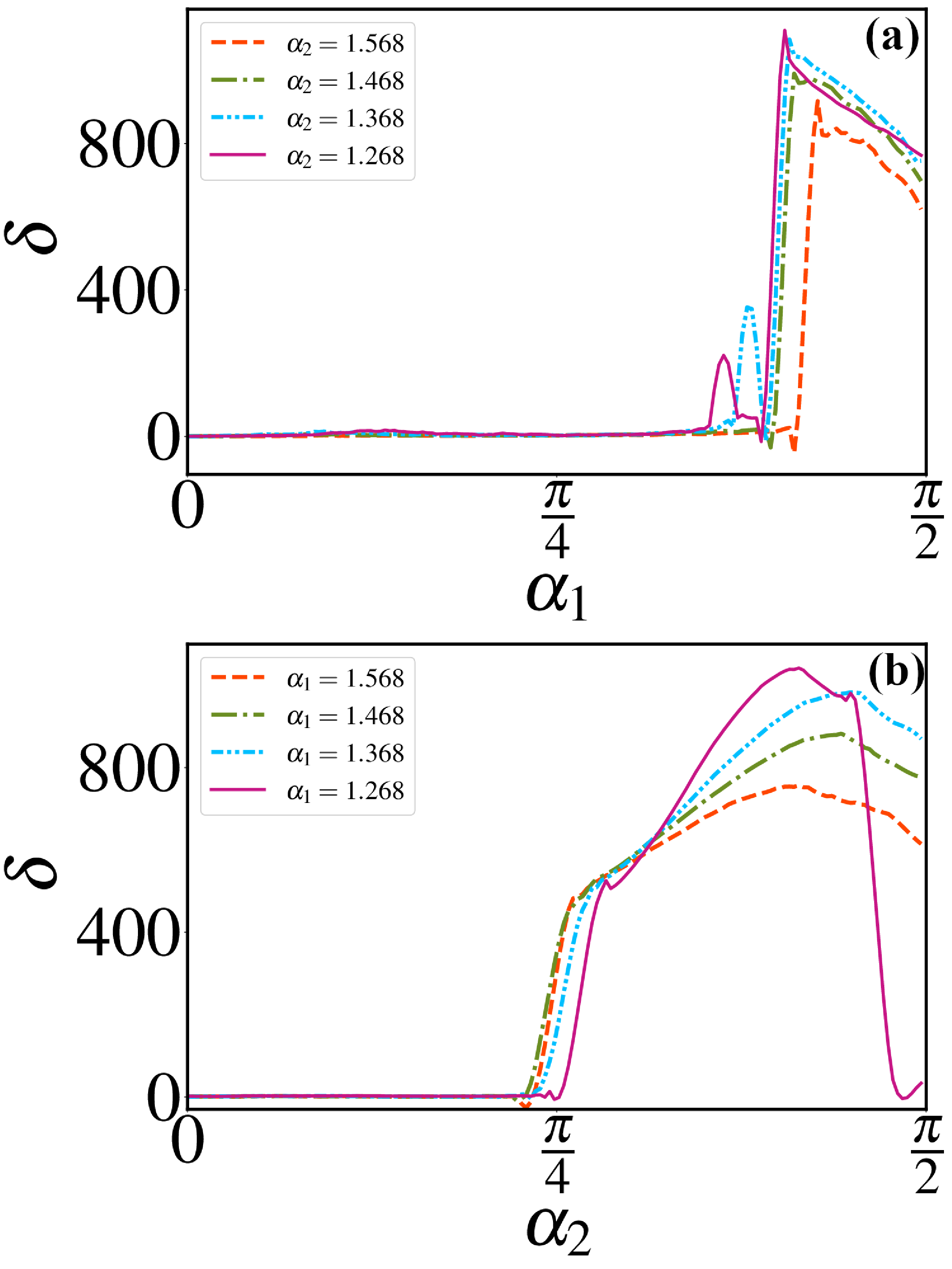}
	\caption{The change in total centroid displacement $\delta$, calculated from Eq. \ref{centroid}, with respect to the frustration parameter $\alpha_{1}$ (with $\alpha_{2}$ fixed) in (a) and with varying $\alpha_{2}$ (with $\alpha_{1}$ fixed) in (b) demonstrates the transition of swarmalators from a non-GTM to a GTM state.
	}
	\label{fig12}
\end{figure}

\begin{figure*}[hbt!]
	\hspace*{0.3cm}
	\includegraphics[width=15cm]{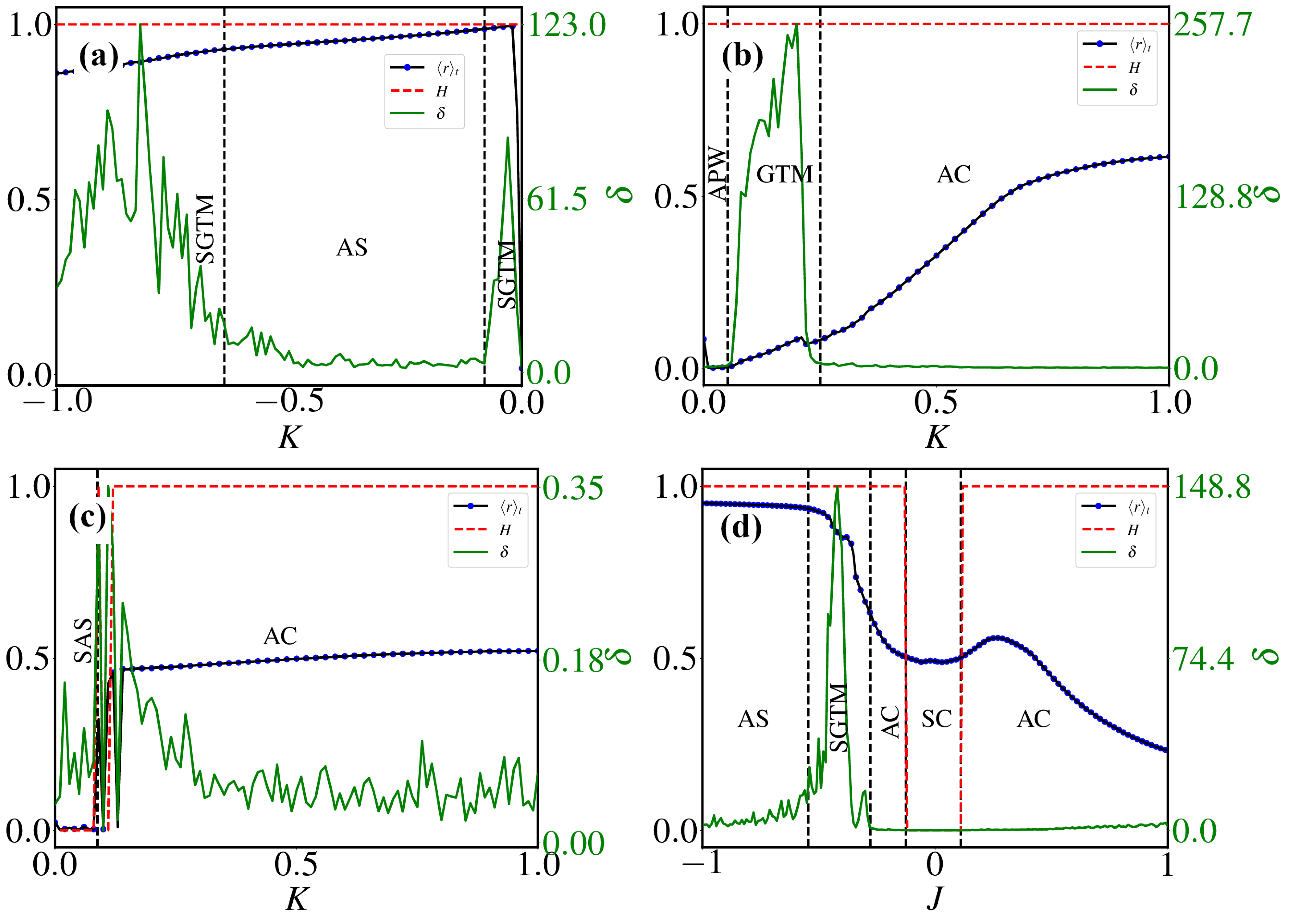}
	\caption{Variation in order parameter values with respect to change in interaction strengths (a) For $J = 0.8$, varying $K$ reveals the existence of SGTM and AS through changes in $\delta$. (b) For $J = 0.8$, varying $K$ indicates the presence of APW, GTM, and AC via $<r>_t$ and $\delta$. (c) For $J = 0.1$, varying $K$ demonstrates the existence of SAS and AC through $H$, $\delta$ and $<r>_t$. (d) varying $J$ for $K = 0.5$, $H$ reveals the presence of SC among other states.}
	\label{fig13}
\end{figure*}

\begin{figure}[hbt!]
	\hspace*{-0.3cm}
	\includegraphics[width=8cm]{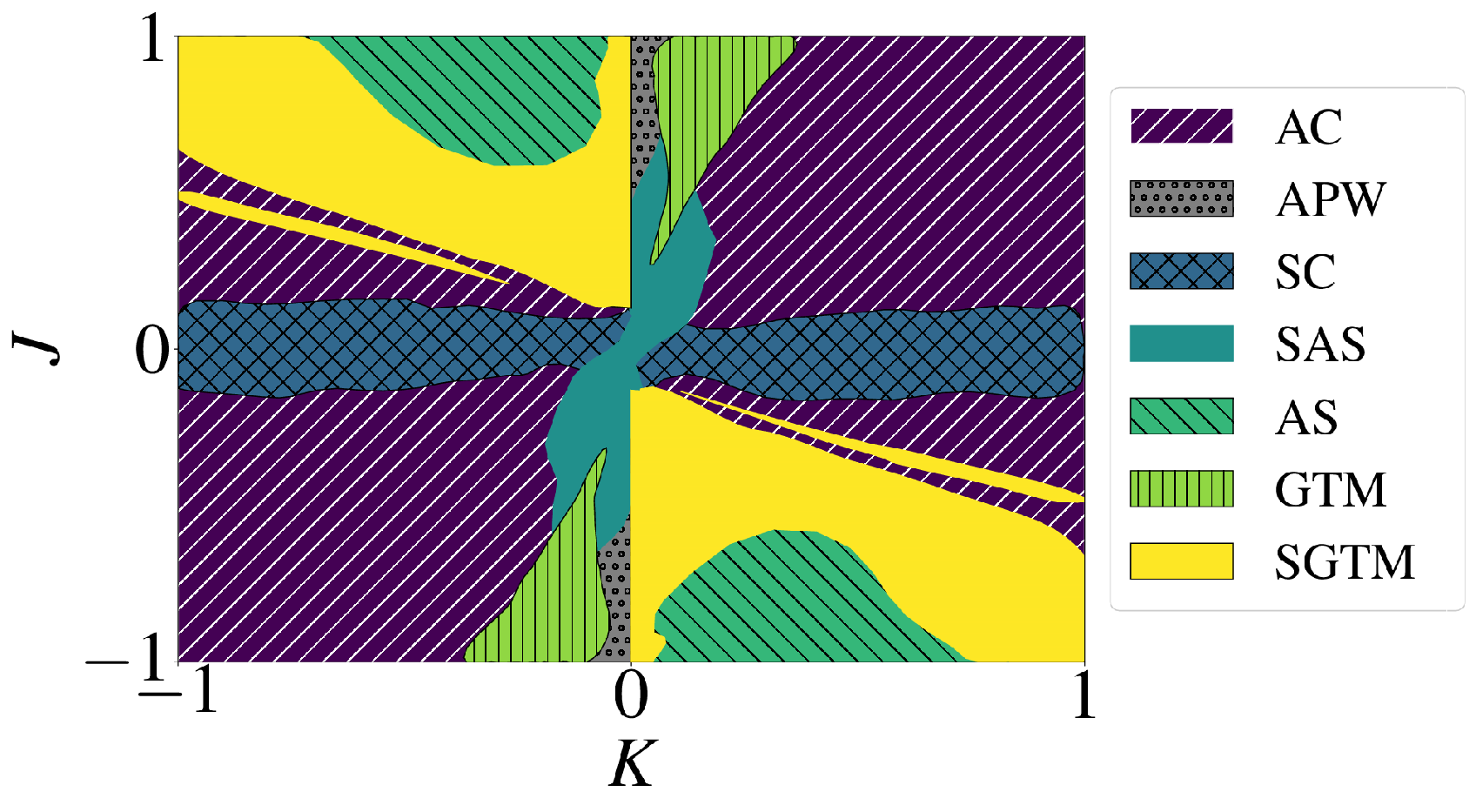}
	\caption{The $J$ vs $K$ plot demonstrates the existence of various dynamical states across different pairs of interaction strengths with $\alpha_{2} = \alpha_{1} = 1.568$.}
	\label{fig14}
\end{figure}

Both GTM and SGTM states emerge due to the inclusion of frustration parameters $\alpha_{1}$ and $\alpha_{2}$. Varying the frustration parameters affects the displacement range and ability of these states to execute global translational motion. Since GTM shows enhanced spatial displacement than SGTM, we can see the impact of frustration parameter in GTM state more clearly. To show the effect of change in frustration parameters in GTM state, we have illustrated the change in total centroid displacement $\delta$ (see equation \ref{centroid}) of GTM with respect to the various values of the frustration parameters $\alpha_{1}$ and $\alpha_{2}$ in Figure \ref{fig12}.  Figure \ref{fig12}(a) shows that GTM occurs only after $\alpha_{1} \approx \frac{3\pi}{8}$, with a delayed onset of GTM as $\alpha_{2}$ increases and minimal variation in GTM range across different $\alpha_{2}$ values. Similarly, the results from Figure \ref{fig12} (b) indicate that swarmalators exhibit GTM only when $\alpha_{2} \geq \frac{\pi}{4}$, regardless of the value of $\alpha_{1}$, highlighting the importance of $\alpha_{2}$ for the initiation of GTM. Furthermore, as $\alpha_{1}$ and $\alpha_{2}$ approaches $\frac{\pi}{2}$ in both the cases, the range of GTM decreases. Therefore, one can confirm that the frustration parameters are crucial and the interplay between interaction strengths $J, K$ and parameters $\alpha_{1}, \alpha_{2}$ gives rise to the states with translational motion. It is worth noting that selecting values of $\alpha_{1}$ and $\alpha_{2}$ near $\frac{\pi}{2}$ enhances the effect of non-stationarity in the system.

\section{Order Parameter-Based State Classification}

In our study, distinct order parameters are examined to investigate the existence of various dynamical states. To differentiate states based on the degree of synchronization, the Kuramoto order parameter is commonly utilized \cite{acebron}. The time-averaged Kuramoto order parameter magnitude $ \langle r \rangle_t $ is defined as:

\begin{align}
	\langle r \rangle_t &= \frac{1}{T} \int_{0}^{T} \left| \frac{1}{N}\sum_{j=1}^{N} e^{\mathrm{i} \theta_{j}(t)} \right| dt
\end{align}

Here, $\theta_{j}$ is the phase of the $j^{th}$ swarmalator. States such as SAS, GTM, and APW, which lack global phase synchronization, yield values of $ \langle r \rangle_t \approx 0 $. In contrast, chimera states like AC and SC exhibit values in the range $ 0 < \langle r \rangle_t < 1 $, while AS and SGTM states, characterized by global synchronization, have values of $ \langle r \rangle_t \approx 1 $.

Furthermore, states involving translational motion, such as GTM and SGTM, can be characterized by calculating the maximum displacement of the centroid $(c_{x(t)}, c_{y(t)})$ of the swarmalators at time $ t $, expressed as follows:

\begin{align}
	(c_{x(t)}, c_{y(t)}) &= \left(\frac{1}{N} \sum_{j=1}^{N} x_j(t), \frac{1}{N} \sum_{j=1}^{N} y_j(t)\right),
\end{align}

\begin{align}
	\delta &= \sqrt{(c_{x(0)} - c_{x(T)})^2 + (c_{y(0)} - c_{y(T)})^2}
	\label{centroid}
\end{align}

\begin{table*}[t]
	\centering
	\begin{tabular}{l@{\hspace{1cm}}c@{\hspace{1cm}}c@{\hspace{1cm}}c}
		\hline\hline
		\textbf{States} & $\mathbf{\langle r \rangle_t}$ & $\mathbf{\delta}$ & $H(\delta)$ \\
		\hline
		Static Asynchronous State (SAS)            & $\sim 0$   & $\sim 0$   & $0$   \\
		Static Chimera (SC)                        & $0 < \langle r \rangle_t < 1$ & $\sim 0$   & $0$   \\
		Active Chimera (AC)                        & $0 < \langle r \rangle_t < 1$ & $\sim 0$   & $1$      \\
		Active Sync (AS)                           & $\sim 1$   & $\sim 0$   & $1$      \\
		Global Translational Motion (GTM)          & $\sim 0$   & $> 0$      & $1$      \\
		Synchronized Global Translational Motion (SGTM) & $\sim 1$   & $> 0$      & $1$      \\
		Active Phase Wave (APW)                    & $\sim 0$   & $\sim 0$   & $1$      \\
		\hline\hline
	\end{tabular}
	\caption{Summary of the order parameters $\langle r \rangle_t$, $\delta$, and $H(\delta)$ for various states.}
	\label{tab:order_parameters}
\end{table*}

where $\delta$ represents the resultant magnitude of the centroid displacement between the initial time and the final time $ T $.

For GTM and SGTM, $ \delta > 0 $, while for all other states, $ \delta = 0 $.

Finally, to distinguish between active and static states, the displacement $ \delta $ is calculated between two arbitrary times $ t_1 $ and $ t_2 $ such that $t_2 > t_1$. However, static states exhibit small displacement value $\delta$ that fluctuates due to the non-stationary behavior in phase and spatial-phase coupling, and active states show large variations in the displacement. To address these variations in the displacement values in the both static and active states, a Heaviside step function is applied, as described by the following expression

\begin{align}
	H(\delta)  =
	\begin{cases} 
		1 & \text{if } \delta > \delta_m \\
		0 & \text{if } \delta \leq \delta_m
	\end{cases}
\end{align}

where $ H(\delta) $ is the transformed value, $ \delta $ is the displacement calculated from Eq. \ref{centroid}, and $\delta_{m}$ is the threshold for the order parameter $H(\delta)$, set at $0.07 + 0.02$. The upper limit of error range $0.02$ arises from slight variations in observed values due to the system's chaotic nature \cite{ansarinasab}. Here the threshold value $\delta_{m}$ is determined by averaging out the maximum spatial displacement of each swarmalator from the population's centroid in the static chimera state for different neighboring parameter values, where each swarmalator exhibits small vibrational motion due to phase non-stationarity. Value of $ \delta $ greater than $ \delta_{m} $ gives $H=1$ representing active state, while $ H=0 $ which corresponds to static states.

The defined order parameters successfully delineate all the identified states, and the corresponding one-parameter plots for different interaction strengths are shown in Figure \ref{fig13}.

In Figure \ref{fig13}(a), for $ J = 0.8 $ and $ -1 < K < 0 $, the translational motion of the SGTM leads to larger displacement values, resulting in higher $\delta$ values. Conversely, for AS, $\delta$ fluctuates around zero due to its non-translational nature. In Figure \ref{fig13}(b), for $ J = 0.8 $ and $ 0 < K < 1 $, the AC state exhibits greater phase synchronization, as evidenced by higher values of the Kuramoto order parameter in the range $ 0 < \langle r \rangle_t < 1.0 $. However, for GTM and APW, $\delta$ serves as the distinguishing factor due to the translational nature of GTM and the non-translational nature of APW. Since all these states are active, $H$ shows unitary value.

In Figure \ref{fig13}(c), for $ J = 0.1 $ and $ 0 < K < 1 $, both $H$ and $\langle r \rangle_t$ yield zero value for SAS, while for AC, these parameters are non-zero, with $\langle r \rangle_t \approx 0.5$, reflecting its chimeric nature. $\delta$ fluctuates around zero as both states are non-translational. In Figure \ref{fig13}(d), for a smoothly varying spatial interaction strength $ J $ from $-1$ to $1$ with fixed $ K = 0.5 $, $H = 0$ indicates the presence of SC around $J=0$, whereas $H=1$ appears elsewhere. The higher $\delta$ values for $ J < 0 $ suggest the existence of SGTM, with $\langle r \rangle_t \approx 1 $ for AS in $ J < 0 $, and $ 0 < \langle r \rangle_t < 1 $ for AC in $ J > 0 $. The Ranges of order parameter values that correspond to each state is illustrated in Table \ref{tab:order_parameters}. 

Figure \ref{fig14} presents a two-parameter plot that sketches the existence of all the collective dynamical states mentioned above for different pairs of $J$ and $K$. These regions are obtained from the order parameters defined above, utilizing the characteristic variations depicted in Figure \ref{fig13}. States such as global translational motion, active phase wave, and static asynchronous state appear in the $(+J, +K)$ and $(-J, -K)$ quadrants. In contrast, synchronized global translational motion and active synchronized states are found in the $(+J, -K)$ and $(-J, +K)$ quadrants. The active chimera state is observed in all quadrants. From Figure \ref{fig14}, it is evident that introducing frustration parameters $\alpha_1$ and $\alpha_2$ alters the landscape of swarmalator dynamics to a greater extent. Notably, we found that the state distribution exhibits inverted symmetry about $J=0$, which is a direct consequence of the system's inherent symmetry under the transformation $(\theta_i, K, J) \rightarrow (-\theta_i, -K, -J)$. For $\alpha_1 = \alpha_2 = \pi/2$, this symmetry ensures that any dynamical state observed at a point $(J, K)$ in the parameter space will have a corresponding state at $(-J, -K)$. In our study, with $\alpha_1$ and $\alpha_2$ taken near $\pi/2$ (specifically, $\alpha_1, \alpha_2 = 1.568$), this symmetry is preserved, resulting in the observed symmetry in the $J-K$ parameter space. This also explains the emergence of phase synchronization behaviors in the $K<0$ regime.

\section{conclusion}
We investigated the effect of the frustration parameters in a two-dimensional swarmalator model. Including the frustration parameter induces non-stationarity in the phases, leading to several intriguing dynamical states, such as chimera and global translational motion (GTM) states. In chimeric states, we identified both static and active chimeras. In the static chimera, the swarmalators form clusters in phases with synchronized and desynchronized properties, where the synchronized population occupies the central region, and the desynchronized population forms a ring around the central region. The active chimera exhibits a breathing-type property, with the swarmalators switching back and forth between synchronized and desynchronized states while oscillating radially in spatial coordinates.

Based on the coupling between phase and spatial movement in swarmalators, non-stationarity in phase dynamics gives rise to complex states such as global translational motion (GTM) and synchronized global translational motion (SGTM). In the GTM state, the entire population of swarmalators moves cohesively in the $xy$ space while simultaneously exhibiting active within-population movements. Compared to SGTM, GTM demonstrates more pronounced motion with higher translational velocity. Additionally, the system exhibits states such as active phase wave (APW) and static asynchronous state (SAS), highlighting the rich dynamical behaviors of swarmalators under the influence of frustration parameters. Our findings will be helpful for significant implications for understanding decentralized control mechanisms in artificial swarm systems such as robotic swarms. By manipulating the frustration parameters, it might be possible to achieve desired collective behaviors in swarms of autonomous robots, which can be applied to various fields such as search and rescue operations \cite{arnold}, environmental monitoring \cite{duarte}, and agricultural automation \cite{albiero}. The future scope of this work lies in the ability to control specific dynamical states like GTM's directional evolution and spatial distribution of chimera states, which might enhance the efficiency and adaptability of swarm systems in complex and dynamic environments.

\section*{Acknowledgements}
		The work of R.G. and V.K.C. forms part of a research project sponsored by ANRF-DST-CRG Project  Grant No. C.R.G./2023/003505. R.G. and V.K.C. thanks DST, New Delhi, for computational facilities under the DST-FIST programme (Grant No. SR/FST/PS-1/2020/135) to the Department of Physics.

 \bibliographystyle{elsarticle-num} 

\begin{thebibliography}{00}

\bibitem{bayani} Bayani, A., Nazarimehr, F., Jafari, S. et al., Nat Commun 15, 4955 (2024). 
	
\bibitem{penn} Penn, Y., Segal, M. and Moses, E., Proceedings of the National Academy of Sciences, 113(12), pp. 3341–3346, (2016).

\bibitem{mccrea} M. McCrea, B. Ermentrout, and J. E. Rubin, Journal of the Royal Society Interface 19, (2022).

\bibitem{pantaleone} J. Pantaleone, American Journal of Physics 70, 992 (2002).

\bibitem{winfree} A. T. Winfree, Journal of Theoretical Biology 16, 15 (1967).

\bibitem{kuramoto} Y. Kuramoto, in Springer eBooks, pp. 420–422, (2005).

\bibitem{guo} Y. Guo et al., International Journal of Electrical Power \& Energy Systems 129, 106804 (2021).

\bibitem{vandermeer} J. Vandermeer et al., Royal Society Open Science 8(3), (2021).

\bibitem{kuramoto_2}Kuramoto, Y., Berlin: Springer, (1984).  

\bibitem{schranz} M. Schranz et al., Frontiers in Robotics and AI 7, (2020).

\bibitem{vicsek} T. Vicsek et al., Physical Review Letters 75, 1226 (1995).

\bibitem{lu} X. Wang, H. Zhao, and L. Li, Applied Sciences 13, 11513 (2023).

\bibitem{puzzo} M. L. R. Puzzo et al., Journal of Physics Condensed Matter 34, 314001 (2022).

\bibitem{christo} H. Christodoulidi et al., in WORLD SCIENTIFIC eBooks, pp. 383–398, (2014).

\bibitem{npetal} N. Phuoc et al., Communications in Physics 23, 121 (2013).

\bibitem{meli} Meli, V.N. et al., Chaos, Solitons \&; Fractals, 177, p. 114278, (2023).

\bibitem{fujiwara} Fujiwara, N., Kurths, J. and Díaz-Guilera, A., Physical Review E, 83(2), 025101, (2011).

\bibitem{sawai} Sawai, S. and Aizawa, Y., Journal of the Physical Society of Japan, 67(8), pp. 2557–2560, (1998).

\bibitem{tanaka} Tanaka, D., Physical Review Letters, 99(13), 134103, (2007). 

\bibitem{iwasa} Iwasa, M. and Tanaka, D., Physics Letters A, 381(36), pp. 3054–3061, (2017).

\bibitem{okeeffe} K. P. O’Keeffe, H. Hong, and S. H. Strogatz, Nature Communications 8, 1504, (2017).

\bibitem{aihara} I. Aihara et al., Scientific Reports 4, 3891, (2014).

\bibitem{riedel} I. H. Riedel, K. Kruse, and J. Howard, Science 309, 300 (2005).

\bibitem{peshkov} A. Peshkov, S. McGaffigan, and A. C. Quillen, Soft Matter 18, 1174 (2022).

\bibitem{hong} Hong, H. et al., Physical Review Research, 5(2), 023105, (2023). 

\bibitem{blum} N. Blum et al., Physical Review. E 109, 014205, (2024). 

\bibitem{sar1} G. K. Sar, D. Ghosh, and K. O’Keeffe, Physical Review. E 107, 024215, (2023).

\bibitem{sar2} G. K. Sar, K. O’Keeffe, and D. Ghosh, Chaos an Interdisciplinary Journal of Nonlinear Science 33, 111103, (2023). 

\bibitem{lizarraga} J. U. F. Lizarraga and M. A. M. De Aguiar, Chaos an Interdisciplinary Journal of Nonlinear Science 30, 053112, (2020).

\bibitem{anwar} M. S. Anwar et al., Communications Physics 7, 59, (2024). 

\bibitem{smith} Smith, L.D., SIAM Journal on Applied Dynamical Systems, 23(2), pp. 1133–1158, (2024).

\bibitem{senthamizhan} Senthamizhan, R., Gopal, R. and Chandrasekar, V.K., Physical Review E, 109(6), 064303, (2024).

\bibitem{sar3} Sar, G.K. et al., New Journal of Physics, 24(4), p. 043004, (2022).

\bibitem{ansarinasab} S. Ansarinasab, F. Nazarimehr, F. Ghassemi, D. Ghosh, and S. Jafari, Appl. Math. Comput. 468, 128508 (2024).

\bibitem{yoon} S. Yoon et al., Physical Review Letters 129, 208002, (2022).

\bibitem{okeeffe2} K. O’Keeffe, S. Ceron, and K. Petersen, Physical Review. E 105, 014211, (2022).

\bibitem{sar4} G. K. Sar, K. O’Keeffe, and D. Ghosh, Chaos an Interdisciplinary Journal of Nonlinear Science 33, 111103, (2023).

\bibitem{sakaguchi} Sakaguchi, H. and Kuramoto, Y., Progress of Theoretical Physics, 76(3), pp. 576–581,  (1986).

\bibitem{lizarraga1} J. U. F. Lizárraga and M. a. M. De Aguiar, Physical Review. E 108, 024212, (2023).

\bibitem{mano} M. Manoranjani, S. Gupta, and V. K. Chandrasekar, Chaos an Interdisciplinary Journal of Nonlinear Science 31, 083130, (2021).

\bibitem{mano1} M. Manoranjani et al., Physical Review. E 105, 034307, (2022).

\bibitem{hsia} C.-H. Hsia et al., Journal of Differential Equations 268, 7897 (2020).

\bibitem{moyal} B. Moyal et al., Physical Review. E 109, 034211, (2024).

\bibitem{marten} E. A. Martens, C. Bick, and M. J. Panaggio, Chaos an Interdisciplinary Journal of Nonlinear Science 26, 094819, (2016).

\bibitem{ha} S.Y. Ha, Y. Kim, and Z. Li, SIAM J. Appl. Dyn. Syst. 13, 466 (2014).

\bibitem{arenas} P. S. Skardal and A. Arenas, Sci. Adv. 1, e1500339 (2015).

\bibitem{sarfati} R. Sarfati and O. Peleg, Science Advances 8, 46, (2022).

\bibitem{kearns} D. B. Kearns, Nature Reviews Microbiology 8, 634 (2010).

\bibitem{kearns1} D. B. Kearns and R. Losick, Molecular Microbiology 49, 581 (2003).

\bibitem{julian} J. Rode et al., Frontiers in Applied Mathematics and Statistics 5, (2019).

\bibitem{bolotov} M. I. Bolotov et al., JETP Letters 106, 393 (2017).

\bibitem{chenko}O. E. Omel’chenko, Journal of Nonlinear Science 32m, 22, (2022).

\bibitem{sheeba} J. H. Sheeba, V. K. Chandrasekar, and M. Lakshmanan, Physical Review E 81, 046203, (2010).

\bibitem{cavagna} A. Cavagna et al., Nature Physics 13, 914 (2017).

\bibitem{acebron} J. A. Acebrón et al., Reviews of Modern Physics 77, 137 (2005).

\bibitem{arnold} R. D. Arnold, H. Yamaguchi, and T. Tanaka, Journal of International Humanitarian Action 3, 18, (2018).

\bibitem{duarte} M. Duarte et al., OCEANS 2016 - Shanghai (2016).

\bibitem{albiero}D. Albiero et al., Computers and Electronics in Agriculture 193, 106608 (2022).


 \end{thebibliography}


\end{document}